# Slit device for FOCCoS – PFS - Subaru

Antonio Cesar de Oliveira[1], James E. Gunn[2], Ligia Souza de Oliveira[3], Marcio Vital de Arruda[1], Lucas Souza Marrara[3], Leandro Henrique dos Santos[1,] Décio Ferreira[1], Jesulino Bispo dos Santos[1], Josimar Aparecido Rosa[1], Flavio Felipe Ribeiro[1], Rodrigo de Paiva Vilaça[1], Orlando Verducci Junior[1], Laerte Sodré Junior[4], Claudia Mendes de Oliveira[4]

1- MCT/LNA-Laboratório Nacional de Astrofísica, Itajubá - MG – Brasil
2- Department of Astrophysical Sciences - Princeton University - USA
3- OIO-Oliveira Instrumentação Óptica LTDA – SP - Brasil
3- IAG/USP – Instituto de Astronomia, Geofísica e Ciências Atmosféricas/ Universidade de São Paulo - SP – Brasil

**ABSTRACT**

The Fiber Optical Cable and Connector System, "FOCCoS", subsystem of the Prime Focus Spectrograph, "PFS", for Subaru telescope, is responsible to feed four spectrographs with a set of optical fibers cables. The light injection for each spectrograph is assured by a convex curved slit with a linear array of 616 optical fibers. In this paper we present a design of a slit that ensures the right direction of the fibers by using masks of micro holes. This kind of mask is made by a technique called electroforming, which is able to produce a nickel plate with holes in a linear sequence. The precision error is around 1-μm in the diameter and 1-μm in the positions of the holes. This nickel plate may be produced with a thickness between 50 and 200 microns, so it may be very flexible. This flexibility allows the mask to be bent into the shape necessary for a curved slit. The concept requires two masks, which we call Front Mask, and Rear Mask, separated by a gap that defines the thickness of the slit. The pitch and the diameter of the holes define the linear geometry of the slit; the curvature of each mask defines the angular geometry of the slit. Obviously, this assembly must be mounted inside a structure rigid and strong enough to be supported inside the spectrograph. This structure must have a CTE optimized to avoid displacement of the fibers or increased FRD of the fibers when the device is submitted to temperatures around 3 degrees Celsius, the temperature of operation of the spectrograph. We have produced two models. Both are mounted inside a very compact Invar case, and both have their front surfaces covered by a dark composite, to reduce stray light. Furthermore, we have conducted experiments with two different internal structures to minimize effects caused by temperature gradients.

This concept has several advantages relative to a design based on Vgrooves, which is the classical option. It is much easier and quicker to assemble, much cheaper, more accurate, easier to adjust; and it also offers the possibility of making a device much more strong, robust and completely miniaturized.

## 1. INTRODUCTION

The Fiber Optical Cable and Connector System (FOCCoS), consists of 3 optical fiber cables that guide light from 2394 positioners to 4 spectrographs.[01] They are called cable A, cable B and cable C. The complete the subsystem, as partitioned by the Subaru PFS project, also includes microlenses, strain relief boxes, fibers plates, multi fiber connectors, and slit devices. The route of FOCCoS goes through the PFI chamber and follows the structure of the telescope to the spectrographs room. Cable A will be installed at the spectrograph side, and will the slit system. This cable is composed of small conduit tubes, containing furcation tubes, which contain groups of optical fibers to be distributed to 4 slit devices, which feed the spectrographs. At the spectrograph room side, cable A is split into 4 branches, each one containing 616 fibers for each spectrograph. Inside each spectrograph, cable A will terminate with a slit device, which is a linear attay of 616 fibers constructed into an arc that matches the curved field of the spectrograph collimator. The physical size of the slit device will be ~140mm long, in which 600 science fibers plus 16 engineering fibers are held with center-to-center spacing of 213,93μm. The proposed optical design for the spectrograph is based on a Schmidt collimator facing a Schmidt camera.[02] This architecture is very robust, well known and documented. It allows for high image quality with only few simple elements (high throughput) at the expense of the central obscuration, which leads to larger optics. Slit devices are part of this central obscuration, so they are subject to size limitations. The size limitation must be balanced against structural and dimensional stability to optimize overall system performance.

This new concept to building slits presents several advantages over previous approaches. It is much easier to assemble in a short time, less expensive, more accurate, and easier to adjust; it also and a strong, robust, and miniature device. This design of slit ensures the direction of the fibers using nickel masks of micro holes. This kind of mask is made by a technique called electroforming, which is able to produce a nickel plate with holes in a linear array. The precision error is around 1μm in the diameter and 1μm in the positions of the holes. This nickel plate may be produced with a thickness between 50 and 200μm so may be very flexible. This flexibility allows inducing the curvature necessary for a curved slit as shown in the Fig. 1.

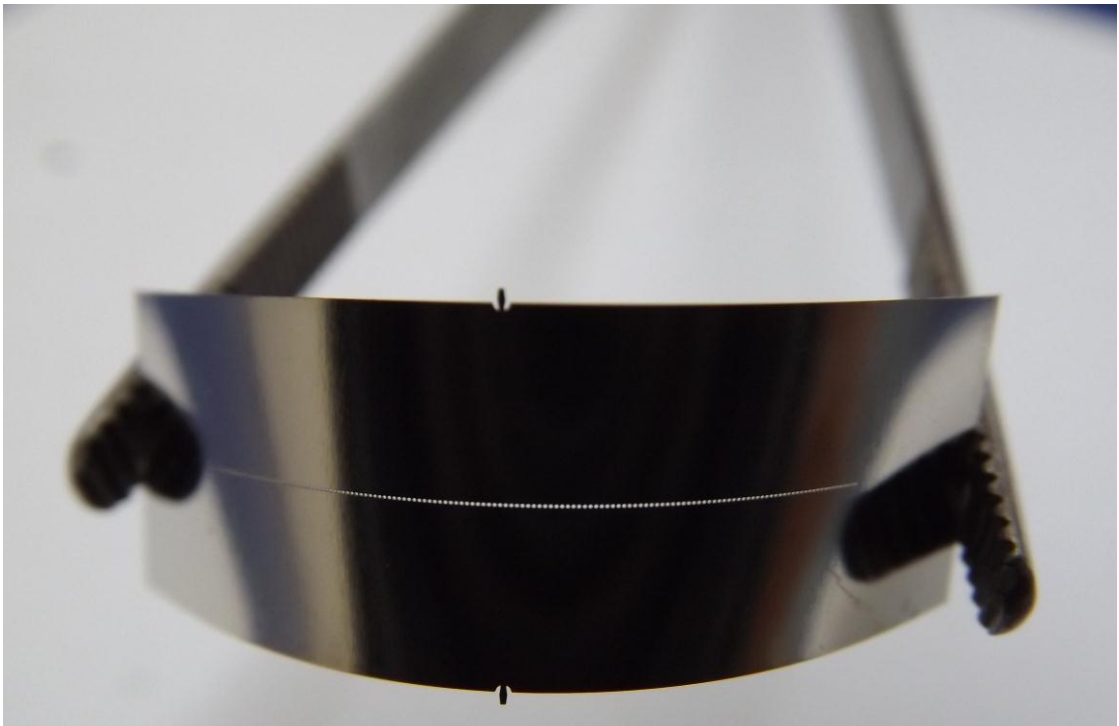

**Figure 1-** Nickel plate mask sample with a sequence of micro holes

The concept in question requires two masks - a Front Mask and a Rear Mask. The two masks are separated by a gap that defines the thickness of the slit, as shown in Figure 2. The pitch and the diameter of the holes define the linear geometry of the slit; the curvature of each mask defines the angular geometry of the slit. Obviously, these parts must be mounted inside some structure that is rigid and strong enough to be supported inside the spectrograph. Furthermore, the entire assembly must have a structural CTE optimized to avoid displacement of the fibers or increase of FRD on the fibers when the device is submitted the low temperature of operation of the spectrograph. (3 deg C)

We describe in this paper two test models submitted to experiments to evaluate effects of FRD in low temperature and displacement of the fibers caused by temperature gradient. The masks with the fibers are mounted inside a very compact invar case, Fig. 3, and the front surface is covered by dark green composite, to reduce light reflections. The dark green composite is obtained from a mixture of EPO-TEK 301-2, ceramics and others materials in nano-particles form. To avoid bubbles and stress, this mixture must be prepared in a separate receptacle inside a vacuum centrifuge machine. The resulting material is more resistant and harder than EPO-TEK 301-2, and it is well suited for the fabrication of optical fiber arrays. An important secondary characteristic is the ease with which it can be polished. This feature is a result of the presence of micro particles, which keep the polished surface very homogeneous during the final polishing procedure.

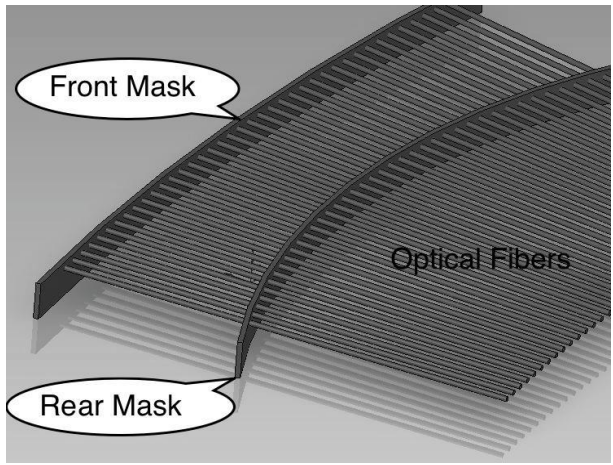

**Figure 2-** Disposition of two masks to define a basic structure of the slit

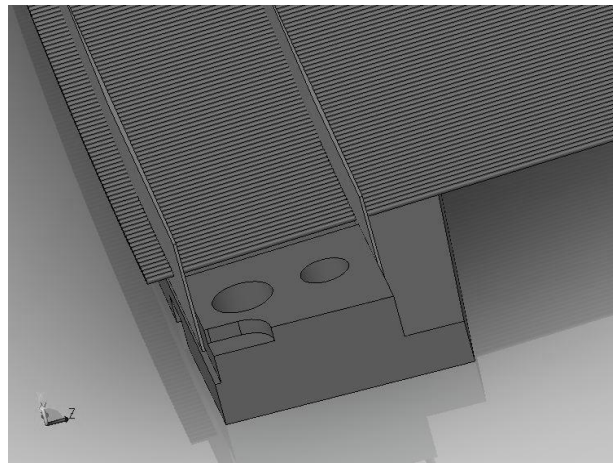

**Figure 3-** INVAR lower base that support the nickel masks and define the slit body

## 2. DESIGN SUGGESTION

The slit optical design for FOOCoS/ PFS follows well-defined requirements for quantities such as the front radius of curvature and the pitch between the consecutive fibers. However, we have constructed prototypes with different dimensions just to use polishing tools from old projects. The results are still applicable to PFS because we are concerned with the assembly procedures and FRD metrology. We have done experimentation with two possible designs. They present small differences that could be significant at the end, in terms of efficiency.

**2.1 Slit device construction**

To obtain a slit that meets all project requirements, all parts of the slit must be measured rigorously. It is very important to develop procedures for mounting the structure. A list of procedures is defined for each step of the construction. We divide assembly into four steps, each one with a basic list of procedures. All steps require using a special jig made in PTFE doped with graphite. This jig is constructed to receive and contain the uncured, liquid dark green composite. After curing, the composite does not adhere to the wall of PTFE, so the slit can be removed of the jig without damage.

The first step starts with fixing the front mask and the rear mask into the INVAR lower base, Fig. 4. The rear mask is glued just following a precision mark, but the front mask must be aligned by a very precise procedure that will be described in the next section. After it is aligned, the front mask is glued in the correct position. Next, the optical fibers are inserted in the holes, through both masks. Finally, the holes in both masks are sealed. This sealing is accomplished in the Front mask by the dark composite; after the composite is dry, the holes is the rear mask are sealed with silicone glue. This procedure fills the gap between the masks, preventing leakage of the by composite in the next step. All procedures of the first step are made on an inclined plane that enables each fiber to be seen as it passes through both masks. The slope also provides resistance to surface tension of the composite on each fiber, preventing it from flowing to the rear mask, which might contaminate the fibers outside the slit. Similarly, the slope also offers resistance to surface tension of the silicone used to seal the holes with fibers inside the rear mask, preventing it from flowing through the slit and contaminating fibers. Obviously it is impossible avoid small effects of surface tension for the composite and silicone, but it is necessary to have some control, so that the composite does not reach the rear mask and the silicone does not exceed the size limit of the structure, Fig. 5.

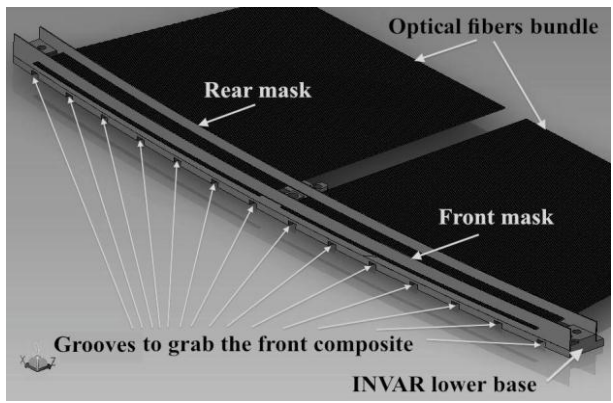
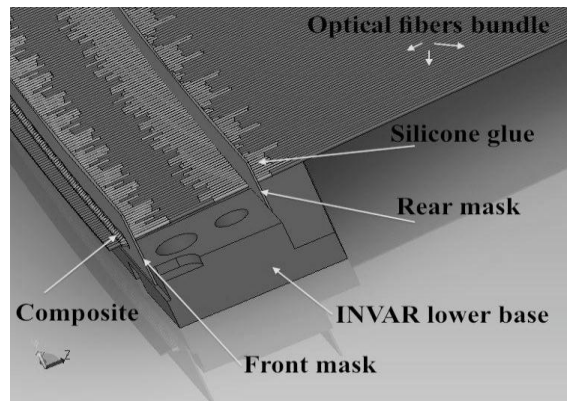

**Figure 4** – First step of the slit assembly that includes; fixing the Rear mask, aligning the Front mask, fixing the Front mask, inserting all optical fibers from cable A, locking holes inside the front mask with dark green composite, and locking holes outside the rear mask with silicone glue.

**Figure 5** – Capillary effect by the composite and the silicone glue on the fibers, during the procedure to seal the holes with fibers. The front mask is sealed with composite and after dried the rear mask is sealed with silicone.

The second step is to screw the INVAR upper base onto the INVAR case, which was constructed in the first step. Guide pins ensure accurate placement. Next, the space between the plates is filled with a potting compound. This potting process ensures a robust, solid slit assembly, protecting optical fibers. We have evaluated two different potting compounds - SYLGARD 184 Silicone Elastomer and Epoteck 301-2 with refractory oxides. ). The objective of this study is to evaluate if each type of nucleation does not increase FRD of the optical fibers. Note that Silicone can contaminate the Epoxy if it is introduced before the Epoxy is completely cured.

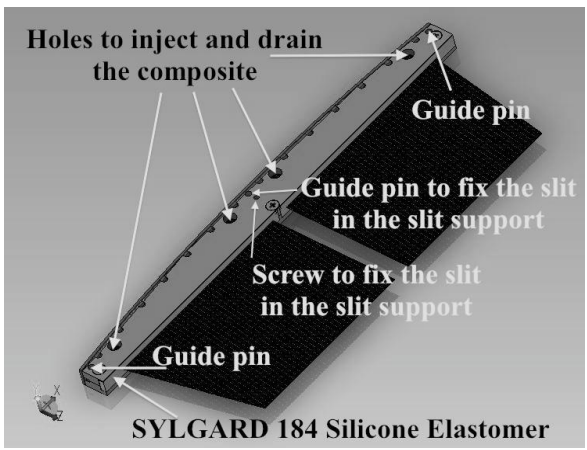
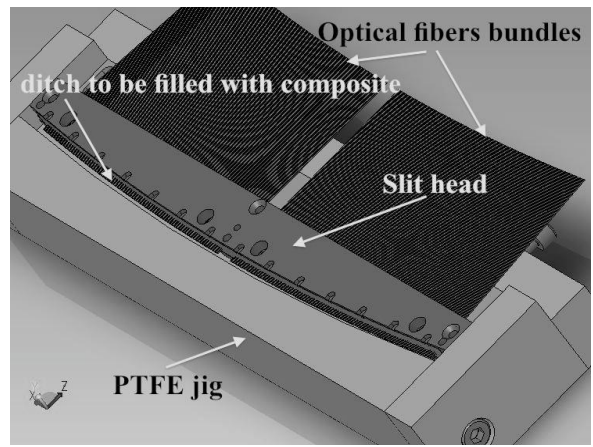

**Figure 6** – Second step of the slit assembly that includes: closing the case with the INVAR upper base, inserting the guide pins to align the lower base against the upper base, screwing the lower base to the upper base, and injecting silicone glue or composite inside.

**Figure 7** – PTFE assembly jig with the slit head to receive the dark green composite that will compose the front surface of the slit. PTFE doped with graphite does not offer adherence for the composite, so it is very easy remove the slit head from the jig when the composite is dried and solidified.

Third step is to make the composite front covering for the slit; this front covering will later be polished and will represent the front surface of the slit. The procedure includes injecting composite in the trench between the front mask and the jig wall, Fig. 7. The composite penetrates the metal grooves giving a structural reinforcement for this covering after curing. This reinforcement is important for two reasons; 1) to avoid detachment of the composite covering from the nickel front mask, and 2) to avoid breaks that might appear due to the high aspect ration of the composite covering. Fig. 8 shows a prototype during the second and third step and Fig. 9 shows the same prototype ready to be polished.

The last step involves polishing composite covering the front surface of the slit. This composite is very attractive for constructing of optical fibre holders because it can be polished with minimum quantities of abrasives. This property is caused by the detachment of the refractory oxide nanoparticles during polishing, which gently reinforces the polishing process, increasing its efficiency. The roughness of the polished surface has been measured to be 0.01 microns. Furthermore, the time for obtaining a polished surface with this quality is about 10 times less than the time required to polish a surface of brass of the same size.

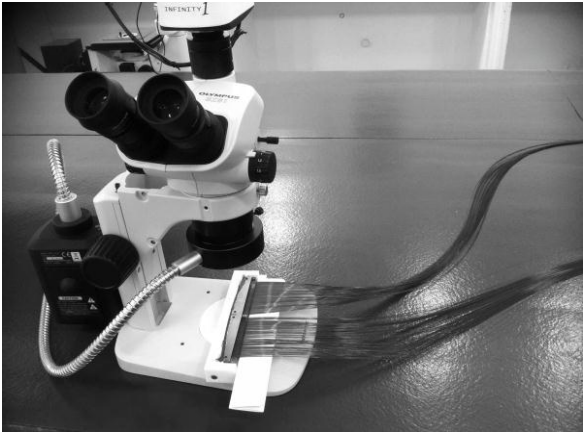

**Figure 8 –** Prototype during steps number 1 and number 2 under the microscope.

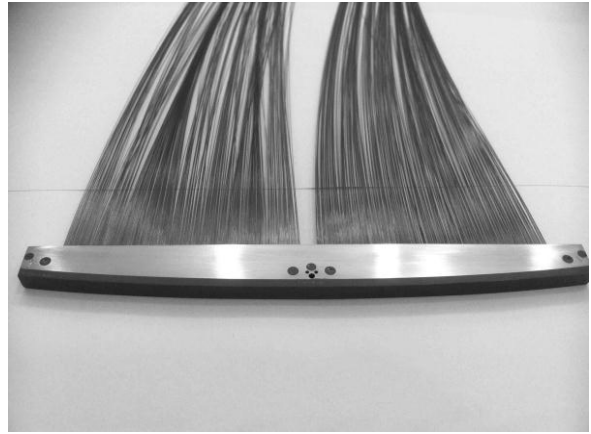

**Figure 9 –** Prototype ready to be polished

The polishing mechanism to be used in curved slits is a system with three axes. This kind of mechanism allows figure-8 motion on a substrate with a cylindrical hollow. Polish quality is checked with a high performance microscope. The polishing process consists initially in the removal of the excess of glue with 30um and 12um emery paper in aluminum oxide film. We have polished small prototypes using a glass paper with good results. However we are developing procedures that use only diamond slurry. Then it is possible to start lapping with 3-micron diamond slurry on a copper plate. A second lapping with 1-micron diamond slurry on a tin-lead plate is needed before the final polishing. The final polishing is made with 0.3um diamond slurry.

**2.2 Mask alignment**

The alignment of masks ensures the aiming fiber so that the projected light from the optical fibers can be routed properly inside the chamber in the spectrograph. The rear mask is fixed using accurate marks printed on the slit body with a laser. Obviously this precision is based in the precise manufacturing of the slit components guaranteed by a suitable metrology. The Rear mask must be displaced until the moment that the bars fill the respective holes, Fig. 10. This operation requires using the microscope and the mask need to receive a temporary adhesive. After alignment, a permanent adhesive is applied to securely hold the Rear mask in the correct position. The alignment of the front mask requires more accuracy and we have developed a mechanism, Fig. 11, that keeps the mask pressed against its mounting surface while still allowing lateral adjustment.

To properly adjust the position of the front mask, two optical fibers are inserted in isometric holes neighboring the slit center. This is an operation done under a Vision System ***Mitutoyo*HYPER UMAP***. Two vectors that form the right angle for the two optical fibers of reference define a set of four points. The points defined by $(X1, Y1)$ and $(X2, Y2)$ are fixed as function of the convergence of the reference vectors, Figs. 12a and 12b. The alignment process resides in to do the adjustment the front mask such that the reference fibers overlapping the reference vectors. This happens when we obtain the correct coordinates for the set points, $(X3, Y3)$ and $(X4, Y4)$.

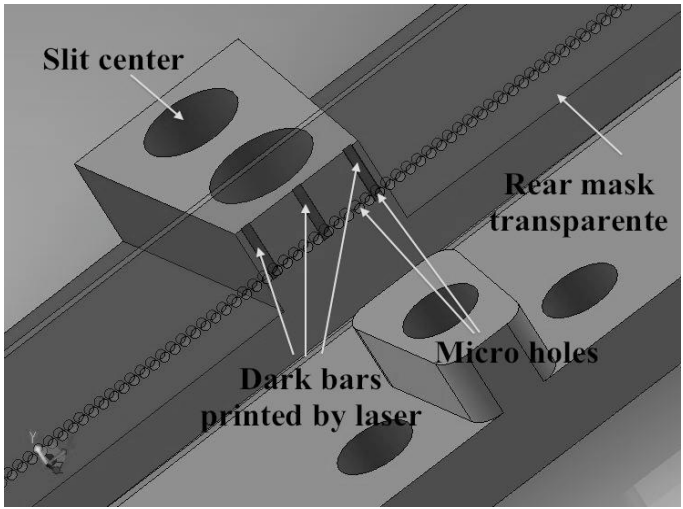
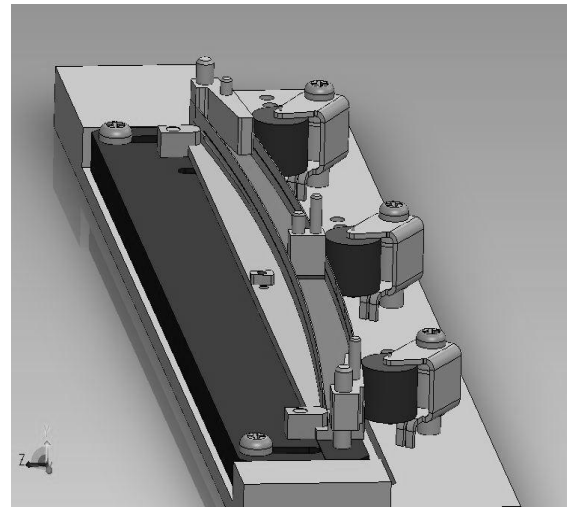

**Figure 10 -** The alignment of the rear mask is set when the position of the mask matches the central holes of the mask with laer-printed black bars on the central support wall. This operation is performed under a microscope.

**Figure 11 –** Device developed to displace the front mask during alignment. Rubber pinch rollers produce constant pressure on the mask, keeping it on its mounting surface while allowing lateral adjustment.

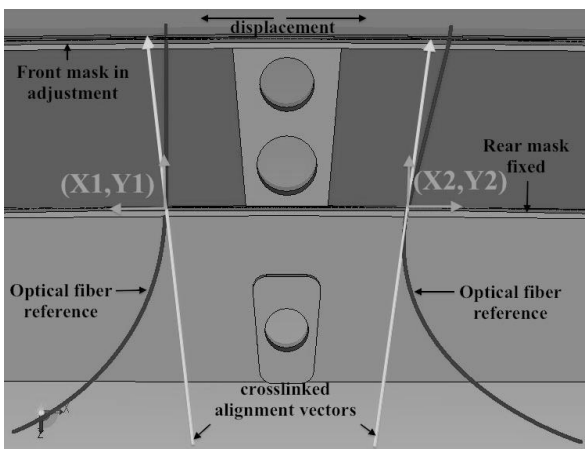
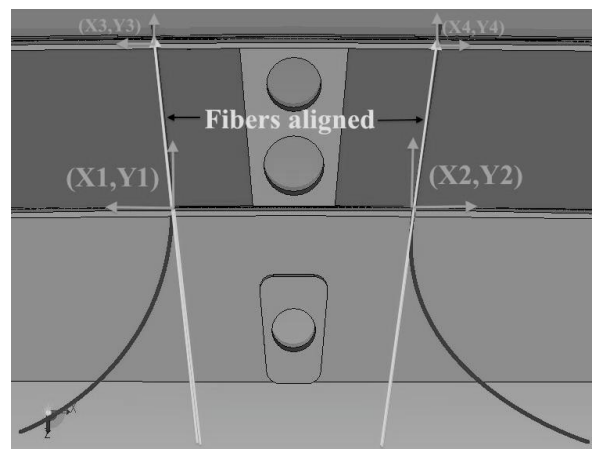

**Figure 12 -** Schematic showing the image of the slit obtained by the Vision System. The front mask in alignment can be moved in such a way that the fibers are displaced until overlapping the reference coordinates (X3,Y3) and (X4,Y4) . In Fig. 12 a, the front mask is misaligned and in Fig. 12 b, the front mask is aligned.

## 3. PROTOTYPING TESTS

Two prototype designs were tested for FRD and thermal stability. They present small differences that could be significant at the end, in terms of efficiency. The geometry of these prototypes does not match PFS requirements, but the test results should allow an informed tradeoff for design options. Figs. 13 and 15, show photos of the prototypes built for testing and analysis. The prototype Model 1, Fig. 14 is potted with silicone (SYLGARD 184 Silicone elastomer) whiles the prototype Model 2, Fig. 16, is potted with composite (Epoteck 301-2 with refractory oxides).

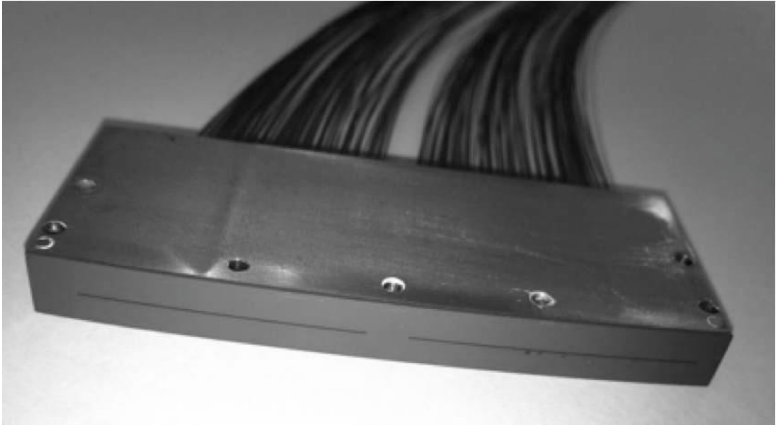
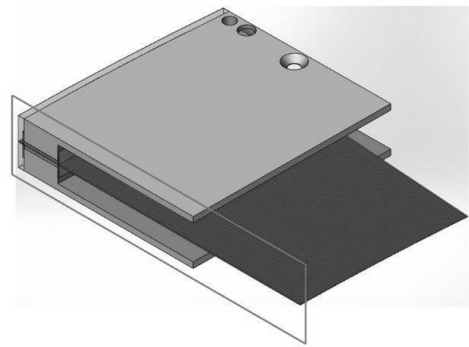
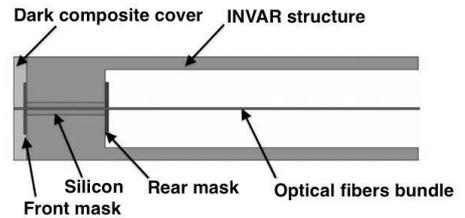

**Figure 13 –** Photo of the prototype Model 2 finished and polished to be tested.
**Figure 14 –** At right, cross section of the prototype Model 2, showing the composite constitution of the core.

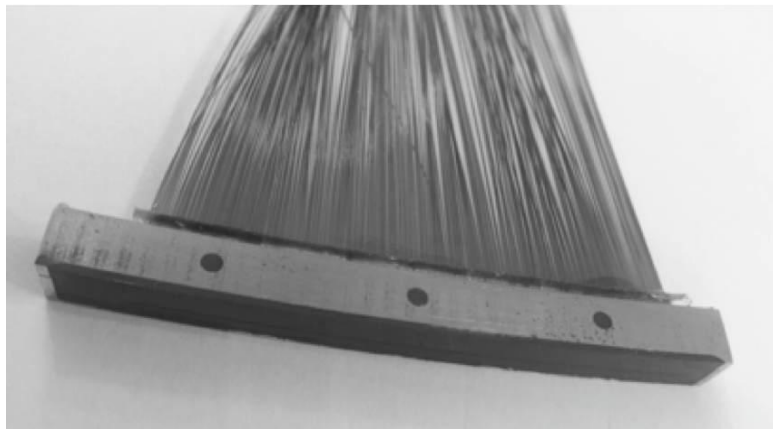
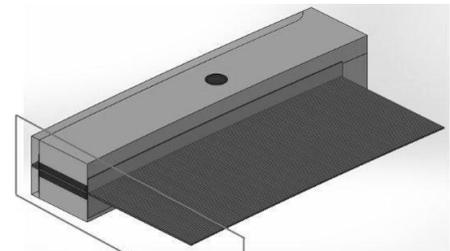
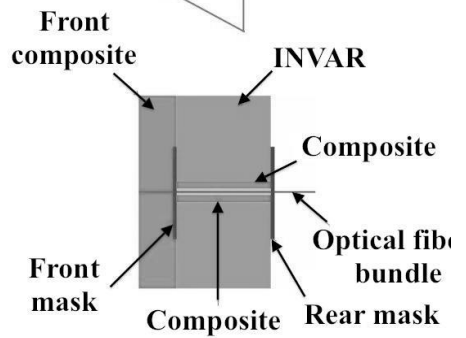

**Figure 15 – Photo** of the prototype Model 1 finished and polished to be tested.
**Figure 16 –** At right, Cross section of the prototype Model 1, showing the silicone constitution of the core.

## 4. EXPERIMENTS AND TESTS AT LOW TEMPERATURE

The goal of these experiments is to measure possible changes on absolute transmission curve of the slit optical fibers caused by temperature changes and the possible displacement of the fibers because the CTE of the slit set. In these experiment the fibers will be tested in two different temperatures: room temperature (19°C ±1°C) and operational temperature (3°C ±1°C)

### 4.1 Submitting the slit prototype at temperature variation

Basically, the slit will be placed between two copper pieces in which one piece is responsible to exchanging heat using a *Peltier* device, controlled by TC-10 Sensym Controller/ with a temperature sensor, and the other piece will be used to read the temperature to over check. The body of the slit sample is mounted on contact with the *Peltier* plate by a cooper bar support. A set of sensors was disposed to check if the temperature along the slit has been in steady state. Silicon grease was used between the Copper plates to guarantee thermal coupling. We have used a Temperature Sensor Module, National (CFP-RTD-122) capable of performing reading from 5 sensors simultaneously.

The complete system is mounted in an X, Y, Z, tilt and rotation stage, Fig.17, so any fiber from the slit device may be positioned to the CCD entrance. To avoid problems with water condensation at low temperature, all structure containing the slit sample, needs to be installed inside a sealed chamber. A positive pressure of dry nitrogen gas is maintained this chamber. With these experimental arrangements it is possible to obtain images of the optical fibers from the slit submitted at low temperatures without water condensation. Our aim was to measure the effect of constriction the optical fiber extremity, caused by the variation in temperature.

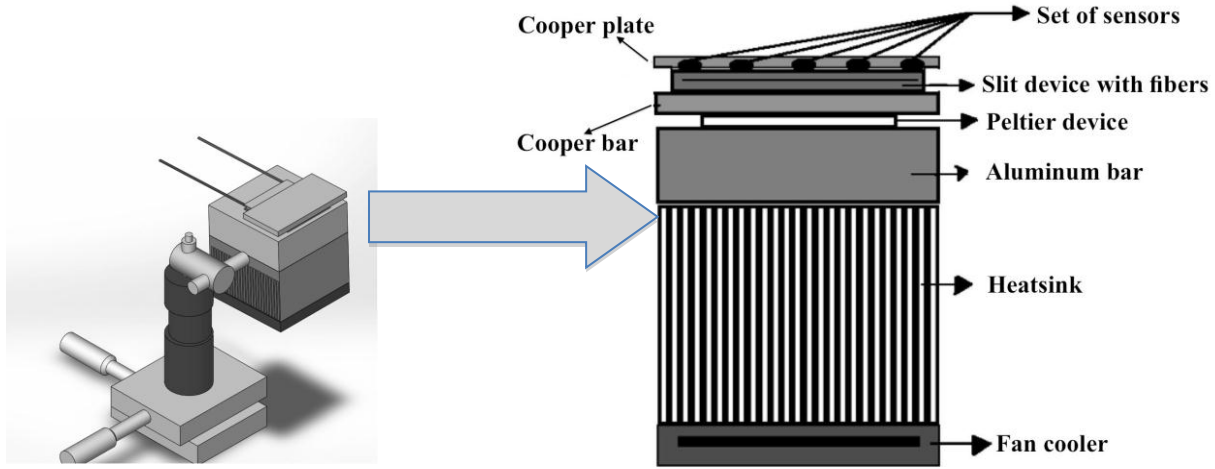

**Figure 17 -** Complete structure to displace in X, Y, Z, adjustment of tilt and rotation to support of the slit coupled a set to control the temperature to do FRD tests at low temperature.

To measure displacement of fiber because the CTE of the slit set, the slit was placed directly in contact with a *Peltier* device which one is responsible to exchanging heat, controlled by TC-10 Sensym Controller with a temperature sensor. Silicon grease is used between the *Peltier* plate and the slit surface to guarantee thermal coupling, Fig.18. To avoid problems with water condensation at low temperature, all structure of including the slit sample, needs to be installed inside a house container with a double glass windows, constructed with special wood, Fig.19. A positive pressure of nitrogen gas is injected inside this house container. With these experimental arrangements it is possible to obtain images of the optical fibers from the slit submitted to low temperatures without water condensation.

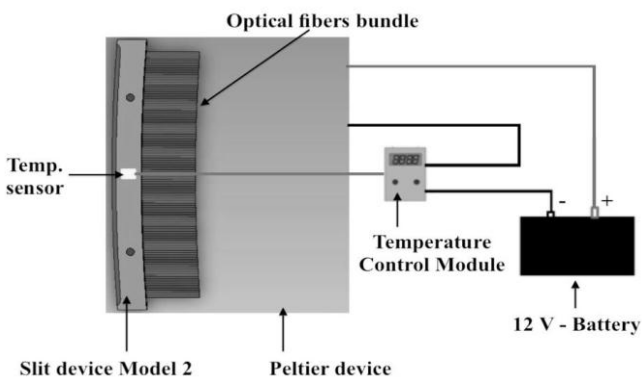

**Figure 18 -** Schematic of the set to cool the slit test and evaluate the displacement of the fibers

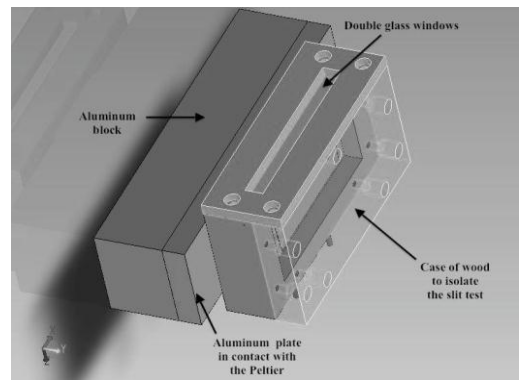

**Figure 19 -** House container to isolate the slit test during the experimentation

Our aim is to measure the possible displacement caused by the CTE of the slit set materials as a consequence of the variation in temperature. The complete system is mounted in a rotation stage, so it is possible to do a correct adjustment of focus for the first and the last optical fiber of the slit test, using the *Mitutoyo HYPER UMAP* Vision System Type2. The inertial variation of temperature by the Sensym Controller is ±1 °C degrees.

**4.2 FRD test**

To measure the Absolute Efficiency we have adapted a method described in Barden.[03] It consists in to illuminate the test fiber with an f/2.8 beam and to compute the ratio between the energy encircled by selected f/# in the fiber output beam and the total energy of the input beam. The experimental apparatus used to measure the FRD is illustrated in Fig. 20.

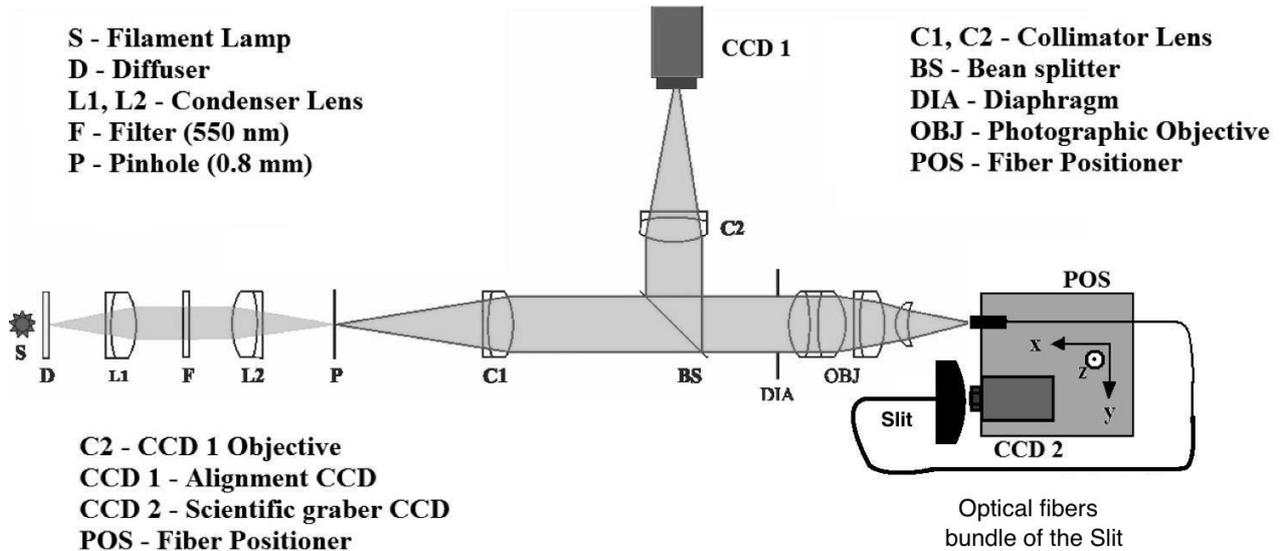

**Figure 20 -** FRD Experimentation – Absolute Transmission in 550 nm ± 10 nm

The pinhole P (0.8 mm) is illuminated by a beam coming from a halogen lamp stabilized S. The uniformity of the beam and its predominant wavelength are obtained using a diffuser D and an interference filter F centered at $\lambda = 550 \pm 10$nm. The main optical system, composed by an achromatic doublet C1 (f = 400 mm) and a photographic lens OBJ (f = 50 mm), produces on measurement plane MP a reduced pinhole image, (0.1 mm). An optical subsystem formed by the lens C2 and a commercial grade camera CCD1 monitor the position and alignment of the optical fiber on MP. This subsystem is connected to optical axis of the main system through the beam splitter BS. A diaphragm DIA placed at the front focal plane of OBJ controls the beam focal ratio. The detector is a scientific grade camera CCD2 placed at the back focal point of an achromatic doublet L4 (f = 12.7 mm). The front focal point of L4 lens is called the focal point of the detector. Initially the system is configured to take reference images. The detector focal point is positioned on the image of the pinhole formed in the measurement plane MP and aligned with the optical axis of the main system by means of a three-axis platform. A reference image is taken in that position, and also a background image. Then the system is configured to take images from the optical fiber. The optical fiber under test has one of its faces positioned over MP and aligned with the optical axis of the main system via a 5-axis platform POS1. The other side of the fiber is positioned and aligned with the detector focal plane using another platform 5-axis POS2. New image is taken as well as a background image. Comparing the reference image with the image data we can determine the amount of energy transmitted by the fiber as well as its spatial distribution of encircled energy.

## 4.3 Software and Calculations

To obtain the absolute transmission of the fiber at a particular Input Focal Ratio, the software takes the concentric annulus centered on the fiber image. This is used to define the efficiency over a range of f-numbers at the exit of the fiber, where each f-number value contains the summation of partial energy emergent from the fiber. Each energy value is calculated by the number of counts within each annulus divided by total number of counts from the image of the spot light inserted at the entrance of the fiber test, with some small defocus, Fig. 21. The limiting focal ratio that can propagate in the tested fiber is approximately F/2.2 taking in account the numerical aperture of this fiber to be 0.22 ± 0.02. Therefore, we have defined F/2.2 to be the outer limit of the external annulus within which all of the light from the test fiber will be collected. The corresponding diameters of the annulus are converted to output focal ratios, multiplying them by the appropriate constant given by the distance between the fiber output end and the detector. To measure possible changes in the Absolute Transmission curve, after obtain a curve A for 19 °C and a curve B for 3 °C, we just need divide B/A. In this case we will obtain:

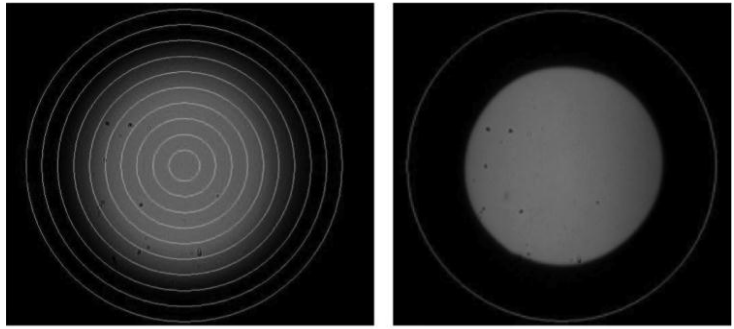

$$E_I \div E_{NA} = \text{Absolute Transmission}$$

**Figure 21:** Mathematical process used to obtain absolute

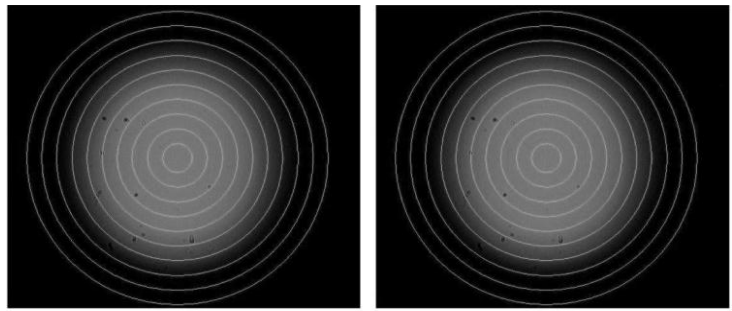

$$E_{NA19\,°C} \div E_{NA3\,°C} = \text{Throughput ratio}$$

**Figure 22-** Description of the methodology adapted to measure any increment of FRD caused by a temperature gradient. The mathematical analysis shows that it is not necessary take any image from the entrance of fiber.

$$\frac{B}{A} = \frac{\frac{EI}{ENA\ 19\ °C}}{\frac{EI}{ENA\ 3\ °C}} = \frac{ENA\ 3\ °C}{ENA\ 19\ °C} \quad (01)$$

However, EI is the same for both measurements less very small variations, so the experimentation may be simplified. Each ratio value is calculated by the number of counts within each annulus from the fiber image at 3°C, divided by number of counts within the corresponding annulus from the fiber image at 19 °C. This give us a direct ratio between the encircled energy, from the fiber at 3 °C with that come from the fiber at 19 °C, Fig. 22.

# 5. RESULTS

## 5.1 – Increment of FRD in the fibers – Slit submitted at temperature gradient

Plots of absolute transmission versus output focal ratio for 2 temperatures were obtained only for the central fiber from the Model 1, Fig. 23. This is a type of experimentation that requires accuracy and stability for a long time necessary to change the position of the CCD among the different conditions. The experimental error estimated for the Absolute Transmission is around 3%.

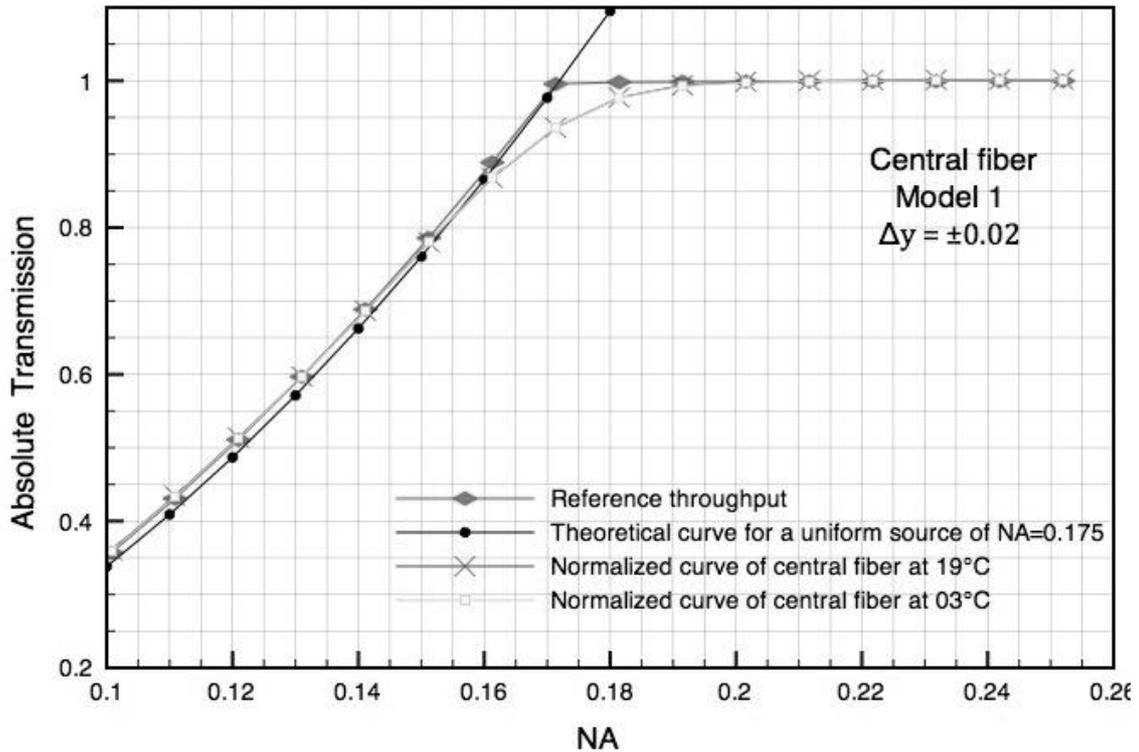

**Figure 23** – Absolute transmission curves for one fiber from the slit Model 1 obtained at 19 °C and 3 °C compared with the theoretical curve for a uniform source of NA 0.175 and the Reference curve from input spot light profile.

A best option to evaluate increments of FRD with subsequent reduction in the Absolute Transmission is just to measure the variation in the light profile from the slit fibers. By this way, we can take images only from the slit fibers avoiding displacements of the structural experimental parts. This is important to reduce the time of the measurement and to be sure about the image position between two different temperatures. Next results were obtained following this procedure, for Model 1 and Model 2. The Fig. 24 shows five throughput ratio curves, obtained from five fibers of the Model 1 and the Fig. 25 shows five throughput ratio curves, obtained from five fibers of the Model 2. Each curve was constructed as average of 3 images taken with 8 minutes time interval, for 19 °C and 03 °C.

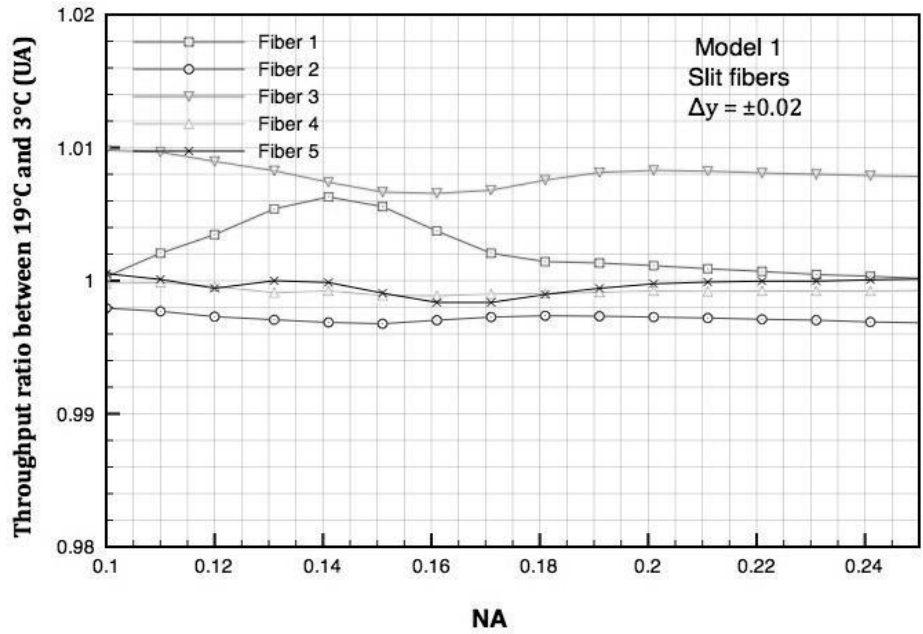

**Figure 24 -** These plots show throughput ratio curves for 19°C and 3°C in the prototype number 1. Average among three different measurements for each fiber.

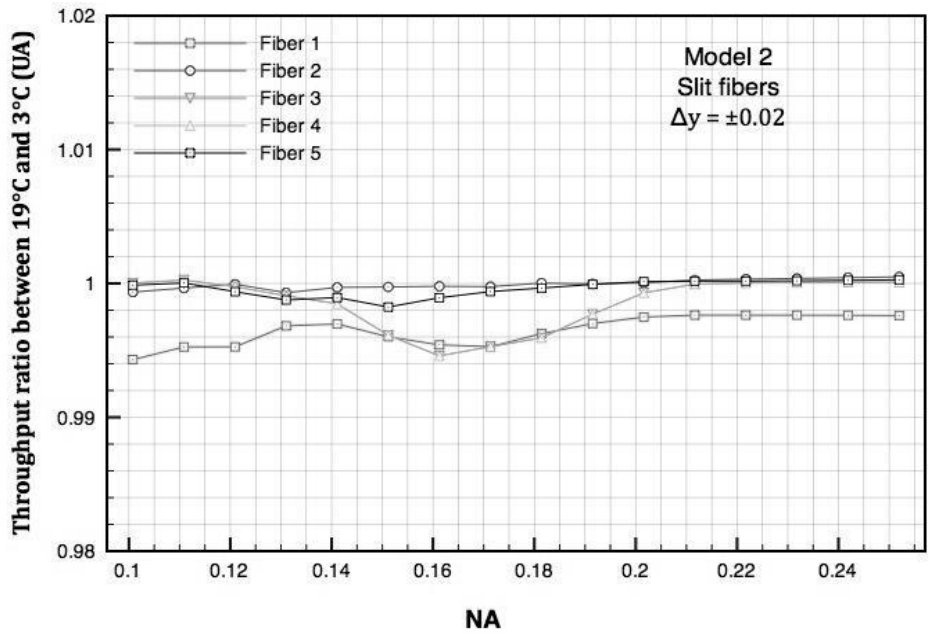

**Figure 25 -** These plots show the throughput ratio curves for 19°C and 3°C in the prototype number 2.

**5.2 – Displacement of fibers – Slit submitted at temperature gradient**

Test made only with prototype Model 2 show us an average distance between the first fiber and the last fiber, border to border of 59.4718 mm. The average displacement between the first and the last fiber, border to border, with a temperature gradient from 20 °C to 03 °C was of 0.0095 mm.

## 6. SUMMARY AND CONCLUSIONS

We described here a new concept to apply in the construction of optical fibers slit device, which ensures the right direction of the fibers by using masks of micro holes. This kind of mask is made by a technique called electroforming, which is able to produce a nickel plate with a linear array of holes. The estimated precision is around 1μm in diameter and 1μm in positions of the holes. The basic project requires two flexible masks, which we call Front Mask and Rear Mask, separated by a gap that define the thickness of the slit. In this context, the pitch and the diameter of the holes define the linear geometry of the slit. The curvature radius of each mask defines the angular geometry of the slit. The internal structure combines INVAR and high performance composite, to ensure good thermal stability inside a range of temperature between 03 °C and 20 °C. Prototypes were constructed to be exposed at tests some of which are presented in this paper. The final propose of this work is develop an efficient slit device for FOCCoS, "Fiber Optical Cable and Connector System", subsystem of PFS, "Prime Focus Spectrograph", that will be installed in the Subaru telescope.

## 7. ACKNOWLEDGMENTS

We gratefully acknowledge support from: Fundação de Amparo a Pesquisa do Estado de São Paulo (FAPESP), Brasil. Laboratório Nacional de Astrofísica, (LNA) e Ministério da Ciência Tecnologia e Inovação, (MCTI), Brasil. We would like also to gratefully Daniel J. Reiley from Caltech-California, for help us with the correction of this paper. Finally we are very grateful to INCT-A (Instituto Nacional de Ciencia e Tecnologia - Astrofisica) to fund our participation in SPIE.